\documentclass[10pt,twocolumn]{article}

\usepackage{relsize}
\usepackage{multicol}
\usepackage{graphicx}
\usepackage{amsmath}
\usepackage{lipsum}
\usepackage{color}
\usepackage{siunitx}
\usepackage{authblk}
\usepackage{subfigure}
\usepackage{amssymb}
\usepackage{wasysym}
\usepackage[margin=2cm]{geometry}

\title{Cooperative effects induced by intruders evolving through a granular medium}
\author[1]{Aymeric Merceron}
\author[1]{Alban Sauret}
\author[1]{Pierre Jop} 
\affil[1]{Surface du Verre et Interfaces, UMR 125 CNRS/Saint-Gobain, 93303 Aubervilliers, France}
\date{ }
\begin{document}

\twocolumn[
    \begin{@twocolumnfalse}
        \maketitle
        \begin{abstract}
            Applying mechanical perturbations in a granular assembly may rearrange the configuration of the particles. However, the spatial propagation of an event is not related to the size of the external perturbation alone. Thus, the characteristic length scale of an event is not well defined. In this study, we trigger rearrangements by driving two intruders through a vertical two-dimensional packing of disks. The amplitude of the rearrangements of the granular assembly appearing around the two evolving intruders is related to their separating distances. We show that there exists a characteristic distance between intruders under which the dynamics of the grains above one intruder is influenced by the other. The size of the intruders has little effect on this characteristic length. Finally, we show that the correlation between the movements of the grains decreases with the distance away from the intruders over a larger length scale. \\
            \medskip\medskip\medskip
        \end{abstract}
    \end{@twocolumnfalse}
]

\section{Introduction}

Granular materials are the most encountered raw materials and are involved in various industrial processes in which the final product may be a homogeneous continuous solid. However, the transformation from an initial granular pile to a final homogeneous phase is not straightforward. {The granular packing may be changed owing to chemical or physical transformations that induce local perturbations \cite{dorbolo2012PRE,gouillart2012insitu}. Therefore, the spatial and temporal extension of perturbations in a granular packing could help to control better or predict the evolution of a pile.}  The typical length scale over which a perturbation propagates in a granular medium is related to both the amplitude of the perturbation and the contact network. Indeed, the local destabilization of an assembly of grains, observed for example when removing some grains, can induce plastic rearrangements close to and far from the perturbation (\textit{e.g.}, \cite{maloney2006plastic}).

The propagation of the perturbation is induced by large spatial heterogeneities of the amplitude of the contact forces in static or flowing granular piles \cite{radjai_force_1996}. The description of these structures has led to the definition of fabric tensors that are related to the shear resistance \cite{voivret_polydisp_2009,azema_fabric_2014}. However, defining relevant length scales in the structure of granular media remains a challenge as they depend on the phenomenon considered (dynamical heterogeneities \cite{dauchot_dynamical_2005}, flow far from a shear band \cite{Nichol2012PRE}). For example, the structural heterogeneities can induce correlated particle motion by releasing mechanical constraints. Also, stable arches can be formed and interrupt the discharge of silos \cite{zuriguel_jamming_2005}. Then, the destabilization of a single grain induces the collapse of the entire structure. The probability of clogging decreases when increasing the width of the orifice, but the existence of a critical width beyond which clogging never occurs is still a matter of debate \cite{janda_jamming_2008, thomas_geometry_2013}. It has also been shown that the jamming probability can be drastically reduced by the nearby presence of a second outlet, whose influence on the flow rate decreases exponentially with the distances between them \cite{kunte_spontaneous_2014}. Similarly, close intruders entering into a granular medium have shown a repulsion effect over a distance of several intruder diameters due to the forced motion of the grains between them \cite{nelson2008projectile,pacheco-vazquez_cooperative_2010}. {Different studies have also considered the spatial propagation of force fluctuations in the granular packing. Such non-local effects allow the granular medium to flow even where the threshold given by the local rheology is apparently not overcome \cite{reddy_evidence_2011,van_hecke_slow_2015}. Such observations can be made by considering} the velocity of the grains in a heap flow that is found to decrease exponentially with the depth \cite{komatsu_creep_2001,kamrin_nonlocal_2012}. As mentioned above, when the jamming transition is approached, the characteristic length scales of correlated motions are increasing \cite{dauchot_dynamical_2005,keys_measurement_2007}. {Most of the previous studies have considered pushed or pulled intruders in a granular packing, as well as the impact of a sphere on a granular bed, to investigate the localization of the flow around the intruder and the cooperative effects \cite{wassgren2003dilute,seguin2008influence,candelier2009creep,clark2012particle,brzinski2013depth,kolb2013rigid,takehara2014high,seguin2016local}. Generally, a jammed state ahead the intruders is created at relatively large velocity and governs the velocity field}.

All these examples exhibit correlated motions in the granular assembly. Long-range correlations can, therefore, impact the dynamic of rearrangement of a {perturbed} granular medium. In this study, we consider the evolution of a two-dimensional granular medium subjected to the synchronized motion of two intruders to study the long-distance effects induced by the transformation of several grains in a granular assembly. More specifically, we focus on the influence of the gap between two intruders on the rearrangement processes. {We go beyond the classical average velocity field or the mean force by characterizing the movement of individual grains.}

This paper is organized as follows. After, a detailed description of experimental and numerical tools, we present the influence of the distance between of the two intruders. We show that the width of the intruder has little effect, and highlight a relevant length scale. We then analyze the amplitude of the perturbations by considering the avalanche size and discuss the existence of typical correlation length scales related to the synchronization of disks movements. We finally conclude by comparing the results obtained in our system with the length scales obtained in different systems previously considered.


\section{Experimental setup and numerical methods}

\begin{figure}
\includegraphics[width=0.48\textwidth]{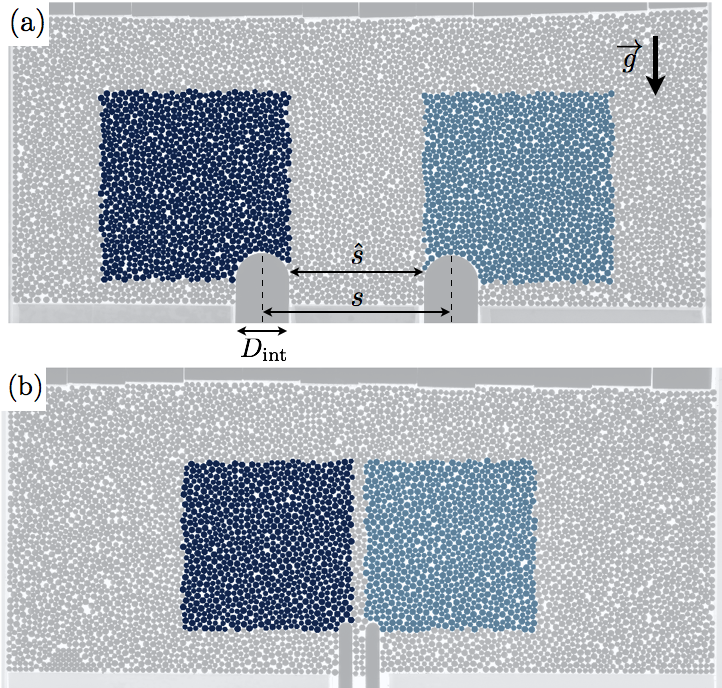}
\caption{Populations of disks analyzed in two different experimental configurations: (a) two intruders of width 10$d_g$ separated by 35$d_g$ and (b) two intruders of width 2.5$d_g$ separated by 5$d_g$.}
\label{fig:pop_analyzed}
\end{figure}

A two-dimensional cell made of two parallel glass plates (30~cm high and 50~cm wide) separated by a gap of 3.1~mm is filled with about 5600 bidisperse stainless steel disks of 4 and 5 mm in diameter to avoid crystallization (proportion 10:7). The experimental setup is an extension of a setup previously used in \cite{merceron_reorganization_2016}. The packing is constrained at the top by confining weights consisting of 12 metallic plates ensuring a uniform pressure over the top of the granular assembly. In the following, all lengths are expressed in terms of diameter of the smallest grains, $d_g$ = 4~mm. Two 3~mm-thick metallic intruders with a semicircular end and of width $D_\mathrm{int}$ are inserted at the bottom of the cell and surrounded by the disks as shown in Fig. \ref{fig:pop_analyzed}(a)-(b). The distance between the centers of the intruders is denoted $s$. The two intruders are then slowly pulled down, out of the cell, at a low and constant speed of 0.05~$\rm{mm.s}^{-1}$ to ensure a quasistatic evolution of the granular packing. {The Froude number in our configurations is $Fr=U/\sqrt{g\,d_g}=2 \times 10^{-4}$, where $U$ is the velocity of the intruder and $d_g$ is the diameter of the grains. In addition, the inertial number is very low, $I=U\,d_g/(W \sqrt{g\,h})=2\times 10^{-5}$, where $W$ is the width of the intruder. These conditions lead to a quasi-static evolution of the system}. The detailed procedure to obtain a reproducible dense initial packing can be found in \cite{merceron_reorganization_2016}. Finally, the pictures of the packing recorded during the experiments are processed with high accuracy, which enables us to obtain uncertainties on the position of each disk smaller than 22 $\mu$m. To obtain reliable statistical data, we repeat each experiment approximatively 40 times for each of the 9 configurations considered in this paper.

Besides, we perform numerical simulations with the discrete element software LMGC90 \cite{LMGC90} to explore a larger number of configurations. Each particle is modeled as a hard disk interacting with its neighbors through the Coulomb friction law, and the dynamics is solved using a non-smooth contact dynamics method \cite{radjai_discrete-element_2011}. The numerical system has the same geometrical parameters as the experiments. The particle-particle and particle-wall dynamical friction coefficients have been measured experimentally and numerically set to 0.13. The disks are sequentially deposited by gravity before computations are performed to finally reach the packing stability. Then, the intruders evolve by successive discontinuous steps of decreasing depths ($\delta z = -1/16 \ d_g$) ending when the system reaches a stable state. The decrease in depth corresponds to the elevation lost by the intruder between two consecutive snapshots. We ensured that this sequential evolution of the numerical intruders does not change the response of the packing under quasi-static conditions. Moreover, the computational time step is set to 10$^{-4}$~s, giving a negligible overlap between disks and ensuring a good accuracy of the simulation. For each of the 28 configurations, the simulations were repeated between 20 and 50 times. We emphasize that the ratio between the disk weight and the overload confining weights is 1.74 larger than in experiments without substantial quantitative changes.

\section{Phenomenological observations}

\begin{figure}
\includegraphics[width=.49\textwidth]{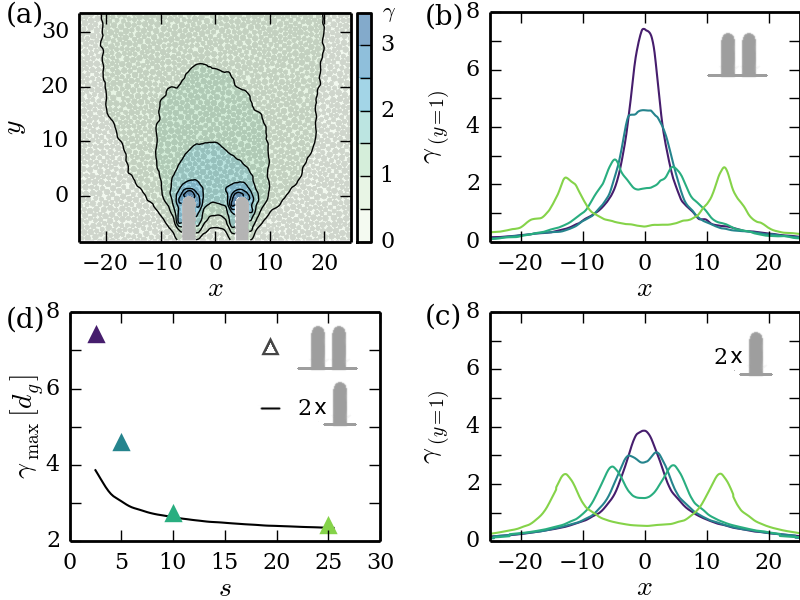}
\caption{(a) Average vertical displacement field $\gamma$ over an experiment for two intruders of width 2.5 $d_g$ and $s=10\,d_g$. Vertical experimental displacement profile (b) recorded just above two intruders of width 2.5$d_g$, $\gamma(y=1)$, and (c) obtained by the addition of the displacement curves measured for a single intruder. (d) Maximal value $\gamma_{\textrm{\scriptsize{max}}}$ for two intruders ($\triangle$) and by addition of two displacement curves measured for a single intruder (continuous line).}
\label{fig:corr_2i}
\end{figure}

We measured the grain displacements between the first and the last images, corresponding to an entire experiment. The resulting displacement field $\gamma$ is shown in Fig.~\ref{fig:corr_2i}(a) for narrow intruders (width 2.5$d_g$). Figure \ref{fig:corr_2i}(b) reports the amplitude of the vertical displacement along the line $y=1$ for four different distances $s$ between the two intruders. For large distances, the curve exhibits two peaks, which merge when the distance $s$ between the intruders decreases. We calculate the superposition of two fields measured for a single intruder to test whether the displacement field for two intruders can be estimated from the displacement field for a single intruder. Figure \ref{fig:corr_2i}(c) exhibits that the sum of the single fields is in good agreement with the displacement above two intruders for large gaps. However, we observe large discrepancies for smaller separation distances between the intruders. In particular, Fig. \ref{fig:corr_2i}(d) shows a fast increase of the amplitude of the displacement peaks in the case of two intruders whose distance is less than about ten particle diameters. Therefore, the influence of a second intruder cannot be described as a simple sum of two displacement fields centered above the intruders. In the following, we analyze the movement of the grains at each time step to better characterize the evolution of the dynamics with the distance of separation.

For both experiments and numerical simulations, we define the avalanche size $N_A$ as the number of disks whose absolute displacement is at least equal to the vertical displacement of the intruder between two time steps \cite{merceron_reorganization_2016}. The avalanche size characterizes the amplitude of a granular reorganization while the response of the granular medium is highly heterogeneous in space and time. For two different widths, we ensure that the normalized distributions of the avalanche size induced by a unique intruder obtained numerically and experimentally are in good agreement.

To measure the influence of one intruder on the neighborhood of the second, we focus our analysis on non-overlapping populations of grains. Instead of measuring the avalanche sizes on the overall population of disks, we analyze only the displacements of disks located in 30 $d_g$-wide squares starting from the inner edge of the intruders as shown in Fig. \ref{fig:pop_analyzed}. The two populations of disks considered are always separated even when the intruders are touching each other and are large enough to get a reliable statistics. The analysis of the avalanche size does not reveal strong differences between the left and right sides of the system, so both sides are considered to investigate the behavior of the packing close to an intruder in the presence of a second one. We should emphasize here that the size of the square areas limits the maximum avalanche size to about 700 disks.    

\begin{figure}
\includegraphics[width=.48\textwidth]{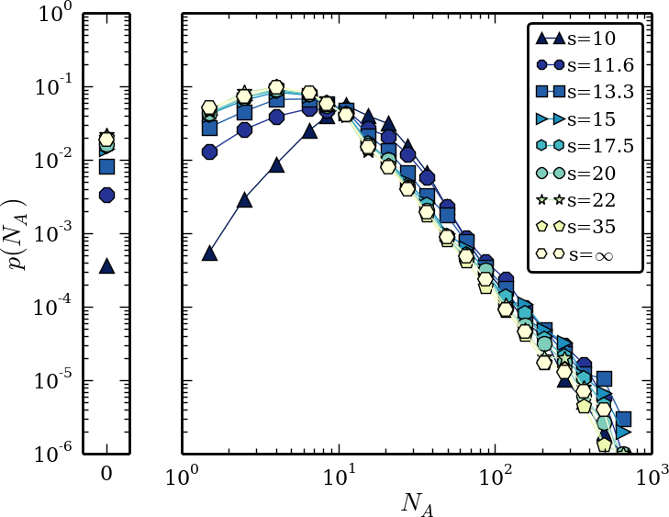}
\caption{Numerical probability density function of avalanche sizes for the 10 $d_g$ wide intruders and for eight separating distances $s$, compared to the distribution for a single 10 $d_g$ wide intruder, {denoted} $s=\infty$ in the legend.}
\label{fig:ShiftDistrib}
\end{figure}

Eight separating distances $s \in$ [10$d_g$, 35$d_g]$ between intruders of width 10$d_g$ have been studied using numerical simulations. {In this figure, the situation $s=\infty$ corresponds to the single intruder case}. The corresponding distributions of avalanche sizes exhibit similar profiles (Fig.~\ref{fig:ShiftDistrib}) to those obtained for a single intruder with a power-law dependence for the largest avalanche sizes \cite{merceron_reorganization_2016}. Such observation is characteristic of processes ruled by scale invariance like snow avalanches \cite{birkeland_power-laws_2002} and earthquakes \cite{gutenberg_magnitude_1956}. Moreover, we observe that these distributions shift toward the distribution obtained for a single intruder when the separation $s$ increases. Hence, the influence of the second intruder spreads over a limited distance and our cell is large enough to avoid boundary effects. In the following section, we focus on the mean value of the avalanche size.

\section{Average avalanche sizes}

\begin{figure}
\includegraphics{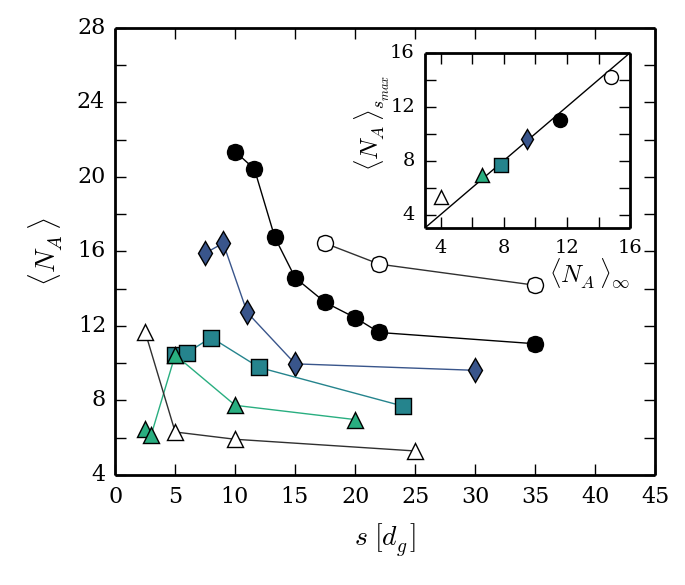}
\caption{Average avalanche size when varying the separating distances $s$. {The open symbols correspond to experimental results, whereas the filled symbols correspond to the results obtained from numerical simulations}. From the lighter to the darker symbols, we have $D_\mathrm{int}/d_g$ = 2.5 ($\vartriangle$), 5 ($\square$), 7.5 ($\lozenge$), 10 ($\ocircle$). Inset: Comparison between the mean avalanche size for a single and two distant intruders.}
\label{fig:mn_avas}
\end{figure}

We now compare the properties of the size distributions when increasing the distance between the two intruders, with the distribution from a single-intruder experiment. For each intruder width, the average avalanche size for the largest separation distance is reported as a function of the mean avalanche size from a unique intruder $\langle N_A \rangle_\infty$ in the inset of Fig. \ref{fig:mn_avas}. The agreement is good, underlying the independence of the intruders, except for the thinnest intruder considered experimentally, which leads to smaller avalanches. In that case, the distribution of avalanche sizes has a flatter tail, whose slope becomes closer to $-2$. The average value of such distribution is not well defined as the distribution is subjected to larger fluctuations because of the finite number of experiments. These results show that we can probe the horizontal range of the perturbations induced by the second intruder. 

We then focus on the evolution of the mean avalanche size when increasing the separating distance $s$ in Fig. \ref{fig:mn_avas}. Starting from the minimal distance, $s_\mathrm{min} = D_\mathrm{int}$, for low separating distances the curves present an increase in the maximum avalanche size before decreasing toward the value obtained for a single intruder at large separation distance. {We shall comment upon the initial increase of the mean avalanche size in the next section.} We observe that the measured mean avalanche size at the contact is close to the one measured for a unique intruder having a width twice the size of the single intruder. However, we also observe small differences; which can be due to the shape of the tip that is not the same in the two experiments. Thus, some grains are trapped between the intruders and introduce a roughness modifying the structure compared to the circular intruder. A second explanation is that the population of grains considered is slightly shifted here since the boundary of the domain lies on the symmetry axis of the two intruders.  Nevertheless, it seems intuitive that the mean avalanche size would initially increase with $s$ because the two intruders with trapped grains in between act as a single, wider intruder, {\textit i.e.}, up to $s - D_\mathrm{int} \approx 1d_g$. However, this trend holds over few more grain diameters. It shows that the behavior above one apparent intruder may enhance the avalanche size and that beyond this optimal distance, we observe a decrease of the influence between the two intruders.

\section{Length scale of cooperative effects}

\begin{figure}
\includegraphics{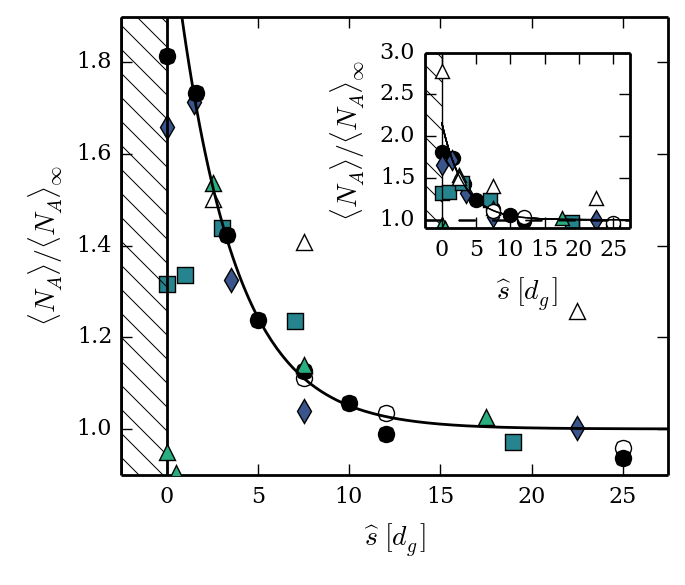}
\caption{Normalized mean avalanche size $\langle N_A \rangle / \langle N_A \rangle_\infty$ as a function of the distance between the inner faces of the two intruders $\widehat{s}$ for the same data as in the figure \ref{fig:mn_avas}. {The open symbols correspond to experimental results, whereas the filled symbols correspond to the results obtained from numerical simulations}. Inset: experimental data for the 2.5$d_g$ wide intruders with the same exponential curve as in the main figure.}
\label{fig:collapse_corr}
\end{figure}

The small value of the optimal distance between two intruders suggests that the relevant length to consider is not the distance between the two tips but rather between the two internal faces of intruders. We thus shift the separating distance using the new variable $\widehat{s} = s - D_\mathrm{int}$. Moreover, considering the asymptotic value for infinite separation, we normalize the mean avalanche size by the corresponding mean avalanche size for a unique intruder $\langle N_A \rangle_\infty$. Figure \ref{fig:collapse_corr} shows the normalized variables. We observe a good collapse of our data for large distances. Yet, the values obtained for low values of $\widehat{s}$ are still dispersed after the normalization. These observations suggest that the mean avalanche size does not scale linearly with the distance between the intruders. Finally, comparing the numerical results with the experiments, we observe the superposition of the data for wide intruders, whereas the curve of the experiment with the narrowest intruders reaches values larger than those predicted numerically (inset of Fig. \ref{fig:collapse_corr}). The master curve decreases exponentially with respect to the separating distance $\widehat{s}>1$ and is well fitted by the expression: 
\begin{equation}
 \frac{\langle N_A \rangle}{\langle N_A \rangle_\infty} - 1 = B\exp\left(-\frac{\widehat{s}}{\lambda}\right),
\end{equation}
where $B$ is a fitting constant and $\lambda$ is a characteristic length scale of the cooperative effects expressed in $d_g$. For our system, $B=1.16$ and $\lambda=3.2 d_g$. The length $\lambda$ is the typical distance beyond which cooperative effects between the two intruders are weak. {This fit is expected not to be valid for $\widehat{s}< 1$, since no grain can be found between the intruders and the system behaves like a single intruder. We observe indeed that the points above $\widehat{s}=1.5$ lay on the master curve.} The good collapse of the data suggests that this length does not depend on the intruder size but on the properties of the granular medium and that the perturbation induced by an intruder spreads over this typical length. We can also note that this length represents the intruder width below which arch formations and jamming events become more frequent inside the granular packing.

{The increase in the mean avalanche size for close intruders corresponds to less frequent jamming events above each intruder and is the consequence of the apparition of correlated motions. The presence of the second intruder modifies the granular structure close to the first intruder. The formation of arches around the tip of one intruder can be destabilized as the force chains evolve above the second intruder. Indeed, when the intruders get closer to each other, stable structures above each intruder share common disks and, as a consequence, the probability of simultaneous collapse is increased. We expect that this effect should be promoted by the narrowness of the intruders as we know that arches are more stable above narrow silo outlets than they are above large silo outlets. This effect is indeed observed to decrease with the width of the intruders (Fig. \ref{fig:collapse_corr}), moreover, for the narrowest intruders, we confirm the existence of long lasting stable arches}. In addition, when the intruders get closer to each other, stable structures above each intruder share common disks and, as a consequence, the probability of simultaneous collapse is increased. Similar phenomena have been found in granular flows through sieves in which the distance between holes is small \cite{chevoir_flow_2007}. In silos with multiple exit orifices and for larger separating distances, the interaction of two arches leads to secondary structures called flying buttresses \cite{mondal_role_2014}. Moreover, the critical spacing between two orifices to observe a mutual influence on the clogging rate lies between 2 and 3$d_g$. In our case, we should emphasize that the evolution of the granular medium is quasi-static. The amplitude of the perturbation is thus not controlled by the speed of the intruder contrary to the creeping motion close to a shear band \cite{reddy_evidence_2011}, in silos with multiple exit orifices \cite{kunte_spontaneous_2014} or in granular bed impacted by multiple projectiles \cite{pacheco-vazquez_cooperative_2010}. In addition, the particle$-$particle friction coefficients should play a role on the typical length found. For example, larger friction coefficients between particles allow the stabilization of arches containing more grains \cite{to_jamming_2001}. We would expect an increase in the characteristic length for larger friction coefficients even if the dissipation of local perturbations were exacerbated. 

\section{Synchronization around intruders}

\begin{figure}
\includegraphics[width=0.48\textwidth]{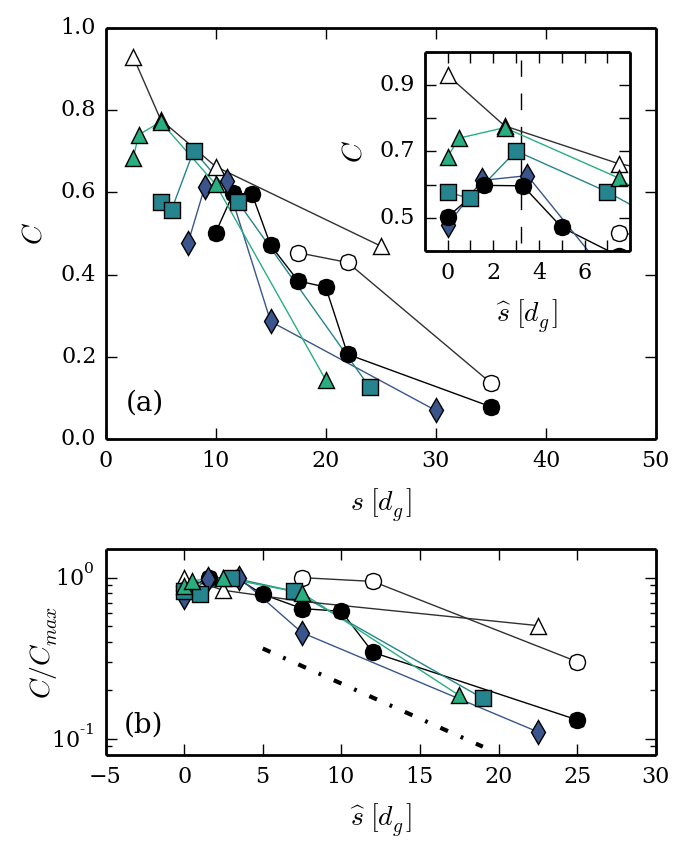}
\caption{(a) Coefficients of correlation between the left and the right intruder as defined in Eq. \ref{eq:autocorr} as a function of the distance $s$ between the intruders. {The open symbols correspond to experimental results, whereas the filled symbols correspond to the results obtained from numerical simulations}. From the lighter to the darker symbols: $D_\mathrm{int}/d_g$ = 2.5 ($\vartriangle$ and $\ocircle$), 5 ($\square$), 7.5 ($\lozenge$), 10 ($\pentagon$ and $\triangledown$). Inset: same data as a function of the shifted distance $\widehat{s}$. The dashed line is located at $\widehat{s}=3.2$$d_g$. (b) Normalized coefficient of correlation as function of $\widehat{s}$. The dash-dotted line is an exponential function of characteristic length of 10$d_g$.}
\label{fig:coor_sync}
\end{figure}

Beyond the amplitude of the mean avalanche size that decreases with the distance between the two intruders, we consider the spatial correlation between events. Indeed, as the intruders get closer and closer to one another, large events located just above an intruder seem to synchronize with the second. To quantify this effect, a correlation function is used to measure how the avalanches occuring on both sides are correlated:
\begin{equation}
 C_i = \frac{\left\langle(N_{A}^{l}(t)\!-\!\langle N_A^{l}\rangle ) (N_{A}^{r}(t)\!-\! \langle N_A^{r} \rangle ) \right\rangle_t}{\sigma_i(N^l_A)\sigma_i(N^r_A)},
\label{eq:autocorr}
\end{equation} 
where $\langle N_A^l\rangle$ stands for the average avalanche size over one experiment $i$ in the vicinity of the left-hand intruder and $\sigma_i(N_A)$ is its standard deviation. The values $C_i$ obtained for each realization in a given configuration are then averaged so that $C = \langle C_i \rangle$. The correlation coefficient is plotted with respect to the separating distances $s$ for different intruder widths in the Fig. \ref{fig:coor_sync}(a). 

The decrease of the value of $C$ at large separating distance is expected, but the evolution of $C$ for close intruders is more surprising. We note that the results from the simulations increase and exhibit a maximum. When using the shifted distance $\widehat{s}$, this maximum is located around the separating distance $\widehat{s}=3d_g$, which corresponds to the characteristic length scale of the decay of the avalanche size with $\widehat{s}$ [inset of Fig. \ref{fig:coor_sync}(a)]. {These unexpected lower correlation coefficients for close intruders can be interpreted by considering the structure. We hypothesize that the structure of the medium is stabilized by the gap between the tips of the two intruders. The few grains between the intruders may act as a stable barrier and screen the propagation of the perturbation through the force network. This stable structure lowers the coupling between each side of the twin intruders.} At large separation distance, we observe a common behavior that follows an exponential decrease as shown in Fig. \ref{fig:coor_sync}(b). The dash-dotted line is proportional to $\exp(-\widehat{s}/10)$. This average slope thus presents a characteristic length which is relatively large, larger than the short length scale revealed by the analysis of the size of the avalanches. This new length scale can be compared to the common lengths observed in granular flows where the typical size of a localized shear band is of the order of ten grain diameters. We measure the decay of the correlation function around a unique intruder and compute the response of two uncorrelated avalanche fields to estimate if this observation can be considered as a simple superposition of the avalanche fields. For a simple superposition, we observe that the decrease of the correlation function is much faster. Our result thus underlines the horizontal reach of the perturbation.
Other configurations documented in the literature exhibit comparable length scales. For example Kunte {\textit et al.} \cite{kunte_spontaneous_2014} observe an exponential increase of the probability of clogging in a silo perforated by a twin aperture whose characteristic separation is of the order of 7-8$d_g$. Pacheco {\textit et al.} measure a repulsive force between penetrating discs that vanishes at an inter-distance of 6$d_g$ \cite{pacheco-vazquez_cooperative_2010}, even if this last situation arises from jamming.

\section{Conclusion}

In this study, we investigated the dynamics of a bidimensional granular packing around two receding intruders numerically and experimentally. We considered the influence of the intruder size as well as the separating distance between them. We observed a sharp increase in the mean avalanche size when the intruders are closer. The associated characteristic length scale is small (3$d_g$) and does not depend on the size of the intruders. This length scale characterizes the typical size of bent force chains and is close to the minimal diameter of an aperture to observe a continuous discharge in a silo. We also computed the correlation function of the avalanches above the two intruders and found a seemingly exponential decrease characterized by a larger length scale of 10$d_g$. This length is linked to the propagation of the perturbation induced by a local rearrangement. At very small separation, we observe a smaller response of the surrounding medium that is the signature of the jammed state of the grains between the tips of the intruders. From our results, we infer that the density of disappearing grains in a {granular} pile will control the intensity of the rearrangements.

\section{Acknowledgments}

The authors acknowledge the support provided by the French ANRT to AM and thank E. Dressaire for a careful reading of the manuscript.

{\small \bibliography{biblio} 
    \bibliographystyle{ieeetr}}

\end{document}